\documentclass{article}%
\usepackage{amsmath}
\usepackage{amssymb}
\usepackage{amsfonts}
\usepackage{graphicx}%
\newtheorem{theorem}{Theorem}

\newtheorem{corollary}[theorem]{Corollary}

\newtheorem{proposition}[theorem]{Proposition}

\newenvironment{proof}[1][Proof]{\noindent\textbf{#1.} }{\ \rule{0.5em}{0.5em}}
\begin{document}

\title{On monotone `metrics' of the classical channel space:non-asymptotic theory}
\author{Keiji Matsumoto\\National Institute of Informatics, Tokyo, Japan}
\maketitle

\begin{abstract}
The aim of the manuscript is to characterize monotone `metric' in the space of
Markov map. Here, `metric' means the square of the norm defined on the tangent
space, and not necessarily induced from an inner product (this property
hereafter will be called inner-product-assumption), different from usual
metric used in differential geometry.

As for metrics in So far, there have been plenty of literatures on the metric
in the space of probability distributions and quantum states. Among them,
Cencov proved the monotone metric in probability distribution space is unique
up to constant multiple, and identical to Fisher information metric. Petz
characterized all the monotone metrics in the quantum state space using
operator mean. As for channels, however, only a little had been known.

In this paper, we impose monotonicity by concatenation of channels before and
after the given channel families, and invariance by tensoring identity
channels. (Notably, we do \textit{not }use the inner-product-assumption. ) To
obtain this result, `resource conversion' technique, which is widely used in
quantum information, is used. We consider distillation from and formation to a
family of channels. Under these axioms, we identify the largest and the
smallest `metrics'. Interestingly, they are \textit{not} induced from any
inner product, i.e., not a metric. Indeed, one can prove that \textit{any}
`metric' satisfying our axioms can \textit{not} be a metric.

This result has some impact on the axiomatic study of the monotone metric in
the space of classical and quantum states, since both conventional theory
relies on the inner-product-assumption. Also, we compute the lower and the
upper bound for some concrete examples.

\end{abstract}

\section{Introduction}

The aim of the manuscript is to characterize monotone `metric' in the space of
Markov map. Here, `metric' means the square of the norm defined on the tangent
space, and not necessarily induced from an inner product, different from usual
metric used in differential geometry.

So far, there have been plenty of literatures on the metric in the space of
probability distributions and quantum states. Cencov, sometime in 1970s,
proved the monotone metric in probability distribution space is unique up to
constant multiple, and identical to Fisher information metric \cite{Cencov}.
He also discussed invariant connections in the same space. Amari and others
independently worked on the same objects, especially from differential
geometrical view points, and applied to number of problems in mathematical
statistics, learning theory, time series analysis, dynamical systems, control
theory, and so on\cite{Amari}\cite{AmariNagaoka}. Quantum mechanical states
are discussed in literatures such as \cite{AmariNagaoka}\cite{Fujiwara}%
\cite{Matsumoto}\cite{Matsumoto}\cite{Petz}. Among them Petz\thinspace
\cite{Petz} characterized all the monotone metrics in the quantum state space
using operator mean.

As for channels, however, only a little had been known. To my knowledge, there
had been no study about axiomatic characterization of distance measures in the
classical or quantum channel space.

In this paper, we impose monotonicity by concatenation of channels before and
after the given channel families, and invariance by tensoring identity
channels. (Notably, we do \textit{not }use the inner-product-assumption. ) To
obtain this result, `resource conversion' technique, which is widely used in
quantum information, is used. We consider distillation from and formation to a
family of channels.

Under these axioms, we identify the largest and the smallest `metric'.
Interestingly, they are \textit{not} induced from any inner product, i.e., not
a metric. Indeed, one can prove that \textit{any} `metric' satisfying our
axioms can \textit{not} be a metric.

In author's opinion, the axiom in this manuscript is reasonable and minimal,
and it is essential that being metric in narrow sense is not required. Hence,
this result has some impact on the axiomatic study of the monotone metric in
the space of classical and quantum states, since both Cencov\thinspace
\cite{Cencov} and Petz\thinspace\cite{Petz} relies on the
inner-product-assumption. Since classical and quantum states can be viewed as
channels with the constant output, it is preferable to dispense with the
inner-product-assumption. This point will be discussed in a separate manuscript.

\section{Notations and conventions}

\begin{itemize}
\item $\mathcal{\Omega}_{\mathrm{in}}$ ($\mathcal{\Omega}_{\mathrm{out}}$)
:the totality of the input (output) alphabet

\item $\mathcal{P}_{\mathrm{in}}$ ($\mathcal{P}_{\mathrm{out}}$) : the
totality of the probability distributions over $\mathcal{\Omega}_{\mathrm{in}%
}$ ($\mathcal{\Omega}_{\mathrm{out}}$). In this paper, the existence of
density with respect to an underlying measure $\mu$ is always assumed. Hence,
$\mathcal{P}_{\mathrm{in}}$ ($\mathcal{P}_{\mathrm{out}}$) is equivalent to
the totality of density functions.

\item $\mathcal{C}$ : the totality of channels which sends an element of
$\mathcal{P}_{\mathrm{in}}$ to an element of $\mathcal{P}_{\mathrm{out}}$

\item $\mathcal{P}_{k}$ : totality of probability mass functions supported on
$\left\{  1,2,\cdots,k\right\}  $

\item $\mathcal{C}_{k,l}$ : totality of the Markov map from $\mathcal{P}_{k}$
to $\mathcal{P}_{l}$

\item $x$,$y$, etc.: an element of $\mathcal{\Omega}_{\mathrm{in}}$
,$\mathcal{\Omega}_{\mathrm{out}}$

\item $X$,$Y$, etc.: random variable taking values in $\mathcal{\Omega
}_{\mathrm{in}}$ ,$\mathcal{\Omega}_{\mathrm{out}}$

\item A probability distribution $p$ is identified with the Markov map which
sends all the input probability distributions to $p$. (Hence represented by a
transition matrix of rank 1.)

\item $\mathcal{T}_{\cdot}\left(  \cdot\right)  $: tangent space

\item $\delta$ etc. : an element of $\mathcal{T}_{p}\left(  \mathcal{P}%
_{\mathrm{in}}\right)  $ etc.

\item $\Delta$ etc. : an element of $\mathcal{T}_{\Phi}\left(  \mathcal{C}%
\right)  $

\item An element $\delta$ of $\mathcal{T}_{p}\left(  \mathcal{P}_{\mathrm{in}%
}\right)  $ etc. is identified with an element $f$ of $L^{1}$ such that $\int
f\mathrm{d}\mu=0$.

\item $g_{p}\left(  \delta\right)  $: \ square of a norm in $\mathcal{T}%
_{p}\left(  \mathcal{P}_{k}\right)  $

\item $G_{\Phi}\left(  \Delta\right)  $: square of a norm in $\mathcal{T}%
_{\Phi}\left(  \mathcal{C}_{k,l}\right)  $

\item $J_{p}\left(  \delta\right)  $ : classical Fisher information

\item The local data at $p$: the pair $\left\{  p,\delta\right\}  $.

\item The local data at $\Phi$ : the pair $\left\{  \Phi,\Delta\right\}  $.

\item $\Phi\left(  \cdot|x\right)  \in\mathcal{P}_{\mathrm{out}}$ : the
distribution of the output alphabet when the input is $x$

\item $\Delta\left(  \cdot|x\right)  \in\mathcal{T}_{p}\left(  \mathcal{P}%
_{\mathrm{out}}\right)  $ is defined as the infinitesimal increment of above

\item $\mathbf{I}$: identity
\end{itemize}

\section{Axioms}

\begin{description}
\item[(M1)] $G_{\Phi}\left(  \Delta\right)  \geq G_{\Phi\circ\Psi}\left(
\Delta\circ\Psi\right)  $

\item[(M2)] $G_{\Phi}\left(  \Delta\right)  \geq G_{\Psi\circ\Phi}\left(
\Psi\circ\Delta\right)  $

\item[(E)] $G_{\Phi\otimes\mathbf{I}}\left(  \Delta\otimes\mathbf{I}\right)
=G_{\Phi}\left(  \Delta\right)  $

\item[(N)] $G_{p}\left(  \delta\right)  =g_{p}\left(  \delta\right)  $
\end{description}

\section{Programming or simulation of channel families}

Suppose we have to fabricate a channel $\Phi_{\theta}$, which is drawn from a
family $\left\{  \Phi_{\theta}\right\}  $, without knowing the value of
$\theta$ but with a probability distribution $q_{\theta\text{ }}$or a channel
$\Psi_{\theta}$, drawn from a family $\left\{  q_{\theta}\right\}  $ or
$\left\{  \Psi_{\theta}\right\}  $. More specifically, we need a channel
$\Lambda$ with
\begin{equation}
\Phi_{\theta}=\Lambda\circ\left(  \mathbf{I}\otimes q_{\theta}\right)
,\,\label{simulation-1}%
\end{equation}
or channels $\Lambda_{a}$ and $\Lambda_{b}$ with
\begin{equation}
\Phi_{\theta}=\Lambda_{b}\circ\left(  \Psi_{\theta}\otimes\mathbf{I}\right)
\circ\Lambda_{a}.\label{simulation-2}%
\end{equation}
Here, note that $\Lambda$, $\Lambda_{a}$, and $\Lambda_{b}$ should not vary
with the parameter $\theta$. Note also that the former is a special case of
the latter. Also, giving the value of $\theta$ with infinite precision
corresponds to the case of $q_{\theta}=\delta\left(  x-\theta\right)  $.

Differentiating the both ends of (\ref{simulation-1}) and (\ref{simulation-2}%
), and letting $\Phi_{\theta}=\Phi$, $q_{\theta}=q$, and $\Psi_{\theta}=\Psi$,
we obtain%

\begin{equation}
\Delta=\Lambda\circ\left(  \mathbf{I}\otimes\delta\right)  ,\text{
}\label{simulation-tangent-1}%
\end{equation}
and
\begin{equation}
\Delta=\Lambda_{b}\circ\left(  \Delta^{\prime}\otimes\mathbf{I}\right)
\circ\Lambda_{a},\label{simulation-tangent-2}%
\end{equation}
where $\Delta\in\mathcal{T}_{\Phi}\left(  \mathcal{C}_{k,l}\right)  $,
$\delta\in\mathcal{T}_{q}\left(  \mathcal{P}_{k^{\prime}}\right)  $, and
$\Delta^{\prime}\in\mathcal{T}_{\Psi}\left(  \mathcal{C}_{k^{\prime}%
,l^{\prime}}\right)  $.

In the manuscript, we consider \textit{tangent simulation}, or the operations
satisfying (\ref{simulation-1}) (or (\ref{simulation-2}) ) and
(\ref{simulation-tangent-1}) (or (\ref{simulation-tangent-2}), resp.), at the
point $\Phi_{\theta}=\Phi$ only. Especially, we are interested in point
simulation of the 1-dimensional subfamily. Note that simulation of $\left\{
\Phi,\Delta\right\}  $ is equivalent to the one of the channel family
$\left\{  \Phi_{\theta+t}=\Phi+t\Delta\right\}  _{t}$.

\section{Relation between $g$ and $G$}

In this section, we study norms with (M1), (M2), (E), and (N).

\begin{theorem}
Suppose (M1) and (N) hold. Then, $\ $%
\[
G_{\Phi}\left(  \Delta\right)  \geq G_{\Phi}^{\min}\left(  \Delta\right)
:=\sup_{p\in\mathcal{P}_{\mathrm{in}}}g_{\Phi\left(  p\right)  }\left(
\Delta\left(  p\right)  \right)  =\max_{x\in\Omega_{\mathrm{in}}}%
g_{\Phi\left(  \cdot|x\right)  }\left(  \Delta\left(  \cdot|x\right)  \right)
.
\]
Also, $G_{\Phi}^{\min}\left(  \Delta\right)  $ satisfies (M1), (M2), (E), and (N).
\end{theorem}

\begin{proof}%
\[
G_{\Phi}\left(  \Delta\right)  =G_{\Phi}\left(  \Delta\right)  \geq
G_{\Phi\circ p}\left(  \Delta\circ p\right)  =g_{\Phi\left(  p\right)
}\left(  \Delta\left(  p\right)  \right)  .
\]
The last identity is trivial. Obviously, $G_{\Phi}^{\min}\left(
\Delta\right)  $ satisfies (M1), (M2) and (N). (E) is seen from the right most
side expression.
\end{proof}

\begin{theorem}
Suppose (M2), (E) and (N) hold. Then
\[
G_{\Phi}\left(  \Delta\right)  \leq G_{\Phi}^{\max}\left(  \Delta\right)
:=\inf_{\Lambda,q,\delta}\left\{  g_{q}\left(  \delta\right)  ;\,\,\Lambda
\circ\left(  \mathbf{I}\otimes q\right)  =\Phi,\,\Lambda\circ\left(
\mathbf{I}\otimes\delta\right)  =\Delta\text{ }\right\}  .
\]
Also, $G_{\Phi}^{\max}\left(  \Delta\right)  $ satisfies (M1), (M2), (E), and (N).
\end{theorem}

\begin{proof}%
\begin{align*}
g_{q}\left(  \delta\right)   &  =G_{q}\left(  \delta\right)  =G_{\mathbf{I}%
\otimes q}\left(  \mathbf{I}\otimes\delta\right)  \geq G_{\Lambda\circ\left(
\mathbf{I}\otimes q\right)  }\left(  \Lambda\circ\left(  \mathbf{I}%
\otimes\delta\right)  \right)  \\
&  =G_{\Phi}\left(  \Delta\right)  .
\end{align*}
So we have the inequality. That $G_{\Phi}^{\max}\left(  \Delta\right)  $
satisfies (M1), (M2), (E), and (N) is trivial.
\end{proof}

\begin{corollary}
\label{cor:gmax>gmin}%
\[
G_{\Phi}^{\max}\left(  \Delta\right)  \geq G_{\Phi}^{\min}\left(
\Delta\right)  .
\]

\end{corollary}

Obviously, $G_{\Phi}^{\min}\left(  \Delta\right)  $ and $G_{\Phi}^{\max
}\left(  \Delta\right)  $ are not induced from any metric, i.e., they cannot
be written as $S\left(  \Delta,\Delta\right)  $, where $S$ is a positive real
bilinear form. Indeed, we can show the following theorem:

\begin{theorem}
Suppose (M1), (M2), (E) and (N) hold. For any interior point $\Phi$ of
$\mathcal{C}_{2,2}$, $G_{\Phi}\left(  \Delta\right)  $ cannot written as
$S_{\Phi}\left(  \Delta,\Delta\right)  $, where $S_{\Phi}$ is a positive real
bilinear form.
\end{theorem}

\begin{proof}
Let $\Phi$ be the one which corresponds to the stochastic matrix
\[
\left[
\begin{array}
[c]{cc}%
1-t & s\\
t & 1-s
\end{array}
\right]  .
\]
Also, let
\[
\Delta_{1}:=\left[
\begin{array}
[c]{cc}%
1 & 0\\
-1 & 0
\end{array}
\right]  ,\,\Delta_{2}:=\left[
\begin{array}
[c]{cc}%
0 & 1\\
0 & -1
\end{array}
\right]  .
\]
Since the family $\{\Phi+\theta\Delta_{1}\}_{\theta}$ can be simulated by the
simulation suggested by the decomposition%
\[
\Phi+\theta\Delta_{1}=\left(  1-t+\theta\right)  \left(  \Phi+t\Delta
_{1}\right)  +\left(  t-\theta\right)  \left(  \Phi-\left(  1-t\right)
\Delta_{1}\right)  ,
\]
(M2) and (E), we have $G_{\Phi}\left(  \Delta_{1}\right)  \leq g_{p}\left(
\delta\right)  $, where $p=(1-t,t)$ and $\delta=\left(  1,-1\right)  $. On the
other hand, by chosing input as $\left(  1,0\right)  $, $\left\{  \Phi
,\Delta_{1}\right\}  $ induces $\left\{  p,\delta\right\}  $. Therefore, by
(M1), $G_{\Phi}\left(  \Delta_{1}\right)  \geq g_{p}\left(  \delta\right)  $
and hence
\[
G_{\Phi}\left(  \Delta_{1}\right)  =g_{p}\left(  \delta\right)  .
\]
Similarly, we have
\[
G_{\Phi}\left(  \Delta_{2}\right)  =g_{q}\left(  \delta^{\prime}\right)  ,
\]
where $q=(s,1-s)$ and $\delta^{\prime}=\left(  1,-1\right)  $. Consider the
family $\{\Phi+t\left(  \Delta_{1}+a\Delta_{2}\right)  \}_{t}$. If $\left\vert
a\right\vert <\min\left\{  \frac{1-s}{t},\frac{s}{t},\frac{1-s}{1-t},\frac
{s}{1-t}\right\}  $, this can be generated by the simulation suggested by%
\[
\Phi+t\left(  \Delta_{1}+a\Delta_{2}\right)  =\left(  1-t+\theta\right)
\left(  \Phi+t\Delta_{1}+ta\Delta_{2}\right)  +\left(  t-\theta\right)
\left(  \Phi-\left(  1-t\right)  \Delta_{1}-a\left(  1-t\right)  \Delta
_{2}\right)  .
\]
Therefore, $G_{\Phi}\left(  \Delta_{1}+a\Delta_{2}\right)  \leq g_{p}\left(
\delta\right)  $. On the other hand, by chosing input as $\left(  1,0\right)
$, $\left\{  \Phi,\Delta_{1}+a\Delta_{2}\right\}  $ induces $\left\{
p,\delta\right\}  $. Therefore,
\[
G_{\Phi}\left(  \Delta_{1}+a\Delta_{2}\right)  =g_{p}\left(  \delta\right)  .
\]
On the other hand, if \ $G_{\Phi}\left(  \Delta\right)  =S_{\Phi}\left(
\Delta,\Delta\right)  $ with some linear bilinear form $S_{\Phi}$,
\begin{align*}
G_{\Phi}\left(  \Delta_{1}+a\Delta_{2}\right)   &  =S_{\Phi}\left(  \Delta
_{1}+a\Delta_{2},\Delta_{1}+a\Delta_{2}\right) \\
&  =S_{\Phi}\left(  \Delta_{1},\Delta_{1}\right)  +a^{2}S_{\Phi}\left(
\Delta_{2},\Delta_{2}\right)  +2aS_{\Phi}\left(  \Delta_{1},\Delta_{2}\right)
\\
&  =g_{p}\left(  \delta\right)  +a^{2}g_{q}\left(  \delta^{\prime}\right)
+2aS\left(  \Delta_{1},\Delta_{2}\right)  .
\end{align*}
Hence, it should hold that
\[
a^{2}g_{q}\left(  \delta^{\prime}\right)  +2aS_{\Phi}\left(  \Delta_{1}%
,\Delta_{2}\right)  =0
\]
for any $\left\vert a\right\vert <\min\left\{  \frac{1-s}{t},\frac{s}{t}%
,\frac{1-s}{1-t},\frac{s}{1-t}\right\}  $. Hence, $g_{q}\left(  \delta
^{\prime}\right)  =0$. Since $\delta\neq0$, this is contradiction.
\end{proof}

Observe that the argument parallel with the above proof applies also to
$\mathcal{C}_{k,l}$ ($k$,$l\geq3$). The following property isuseful in
computation of $G^{\max}$.

\begin{proposition}
\label{prop:phi-max-extreme}Let $\left\{  \Upsilon^{\left(  i\right)
}\right\}  _{i=1}^{n}$ be the extreme points of $\mathcal{C}$. Then
\[
G_{\Phi}^{\max}\left(  \Delta\right)  =\min_{q,\delta}g_{q}\left(
\delta\right)
\]
where $q=\left(  q_{1},\cdots,q_{n}\right)  $ is a probability distribution
over $\left\{  \Upsilon^{\left(  i\right)  }\right\}  $ with
\[
\Phi=\sum_{i=1}^{n}q_{i}\Upsilon^{\left(  i\right)  },\,\,
\]
and $\delta=\left(  \delta_{1},\cdots,\delta_{n}\right)  $ satisfies
$\Delta=\sum_{i=1}^{n}\delta_{i}\Upsilon^{\left(  i\right)  }$.
\end{proposition}

\begin{proof}
Consider a simulation suggested by the decomposition
\[
\Phi=\int\Psi\mathrm{d}P\left(  \Psi\right)  ,\,\,\,\Delta=\int\Psi
f\mathrm{d}P\left(  \Psi\right)  ,
\]
where $P$ is a probability measure over $\mathcal{C}$ $\ $and $\int
f\mathrm{d}P\left(  \Psi\right)  =0$. Here the 'program' is $\left\{  P,f\circ
P\right\}  $, where $f\circ P$ is the singed measure defined by $f\circ
P\left(  A\right)  =\int_{A}f\mathrm{d}P\left(  \Psi\right)  $. Letting
$\Psi=\sum_{i=1}^{n}p_{i|\Psi}\Upsilon^{\left(  i\right)  }$, we obtain
another simulation corresponding to the decomposition
\[
\Phi=\sum_{i}q_{i}\Upsilon^{\left(  i\right)  },\,\,\,\Delta=\sum_{i}%
\delta_{i}\Upsilon^{\left(  i\right)  },
\]
where
\[
q_{i}:=\int p_{i|\Psi}\mathrm{d}P\left(  \Psi\right)  ,\,\delta_{i}:=\int
p_{i|\Psi}f\mathrm{d}P\left(  \Psi\right)  .
\]
Here the `program' is the pair $\left\{  q,\,\delta\right\}  $. The following
Markov map sends the pair $\left\{  P,f\circ P\right\}  $ to the pair
$\left\{  q,\,\delta\right\}  $: upon accepting $\Psi$, which is generated
according to the probability measure $P$, generate $\Upsilon^{\left(
i\right)  }$ with the probability $p_{i|\Psi}$. Therefore, by monotonicity,
\[
g_{P}\left(  f\circ P\right)  \geq g_{q}\left(  \delta\right)  ,
\]
which implies the assertion.
\end{proof}

\section{Binary channels $\mathcal{C}_{2,2}$}

In this section, we suppose $g$ is the Fisher information metric.
$\mathcal{C}_{2,2}$ has four extreme points,
\[
\Upsilon^{\left(  1\right)  }:=\left[
\begin{array}
[c]{cc}%
1 & 0\\
0 & 1
\end{array}
\right]  ,\,\Upsilon^{\left(  2\right)  }:=\left[
\begin{array}
[c]{cc}%
0 & 0\\
1 & 1
\end{array}
\right]  ,\,\Upsilon^{\left(  3\right)  }:=\left[
\begin{array}
[c]{cc}%
0 & 1\\
1 & 0
\end{array}
\right]  ,\,\Upsilon^{\left(  4\right)  }:=\left[
\begin{array}
[c]{cc}%
1 & 1\\
0 & 0
\end{array}
\right]  ,\,
\]
and can be parameterized as
\[
\left[
\begin{array}
[c]{cc}%
1-t & s\\
t & 1-s
\end{array}
\right]  .
\]
Hence the space can be viewed as a square. Consider one-parameter subfamily
$\left\{  \Phi_{\theta}\right\}  $ of $\mathcal{C}_{2,2}$, passing through
$\Phi$. Let $\Psi_{A}$ and $\Psi_{B}$ the intersection of the edge of
$\mathcal{C}_{2,2}$ and the tangent line at $\Phi$ with the tangent $\Delta$.
Obviously, $\left\{  \Phi,\text{ }\Delta\right\}  $ can be simulated as a
probabilistic mixture of $\Psi_{A}$ and $\Psi_{B}$. Hence, defining $a$ and
$b$ by $\Delta=a\left(  \Psi_{A}-\Psi_{B}\right)  $ and $\Phi=b\Psi
_{A}+\left(  1-b\right)  \Psi_{B}$,
\[
G_{\Phi}^{\max}\left(  \Delta\right)  \leq\frac{a^{2}}{b}+\frac{a^{2}}{1-b}.
\]

Suppose $\Psi_{A}$ and $\Psi_{B}$ can be discriminated with certainty by
observing the output for a properly chosen input. This occurs if and only if
one of the following is true:
\begin{align*}
\left[  \Psi_{A}\right]  _{11}  &  =1\,\ \&\,\quad\left[  \Psi_{B}\right]
_{01}=1\,,\\
\left[  \Psi_{A}\right]  _{01}  &  =1\,\ \&\,\quad\left[  \Psi_{B}\right]
_{11}=1\,,\\
\left[  \Psi_{A}\right]  _{10}  &  =1\,\ \&\,\quad\left[  \Psi_{B}\right]
_{00}=1\,,\\
\left[  \Psi_{A}\right]  _{00}  &  =1\,\ \&\,\quad\left[  \Psi_{B}\right]
_{10}=1\,.
\end{align*}
In such cases, one can extract the Fisher information of the binary
distribution which is used to mix $\Psi_{A}$ and $\Psi_{B}$. \ Therefore,
\[
G_{\Phi}^{\min}\left(  \Delta\right)  \geq\frac{a^{2}}{b}+\frac{a^{2}}{1-b}.
\]
Hence, due to Corollary\thinspace\ref{cor:gmax>gmin}, we have
\[
G_{\Phi}\left(  \Delta\right)  =G_{\Phi}^{\min}\left(  \Delta\right)
=G_{\Phi}^{\max}\left(  \Delta\right)  =\frac{a^{2}}{b}+\frac{a^{2}}{1-b}.
\]
Especially, if $\Phi=\frac{1}{2}\left[
\begin{array}
[c]{cc}%
1 & 1\\
1 & 1
\end{array}
\right]  $, this is the case for any $\Delta$.

In general, however, the simulation by the mixture of $\Psi_{A}$ and $\Psi
_{B}$ is not optimal. For example, let
\begin{align*}
\Phi &  :=a\Upsilon^{\left(  1\right)  }+b\Upsilon^{\left(  2\right)
}+c\Upsilon^{\left(  3\right)  }=\left(  a-t\right)  \Upsilon^{\left(
1\right)  }+\left(  b+t\right)  \Upsilon^{\left(  2\right)  }+\left(
c-t\right)  \Upsilon^{\left(  3\right)  }+t\Upsilon^{\left(  4\right)  }\\
&  =\left[
\begin{array}
[c]{cc}%
a & c\\
b+c & a+b
\end{array}
\right]  =\left[
\begin{array}
[c]{cc}%
a & c\\
1-a & 1-c
\end{array}
\right]  ,\text{ }\\
\quad\Delta &  :=\left[
\begin{array}
[c]{cc}%
-1 & 1\\
1 & -1
\end{array}
\right]  =\Upsilon^{\left(  3\right)  }-\Upsilon^{\left(  1\right)  }=\left(
1-s\right)  \Upsilon^{\left(  3\right)  }+s\left(  \Upsilon^{\left(  2\right)
}+\Upsilon^{\left(  4\right)  }-\Upsilon^{\left(  1\right)  }\right)
-\Upsilon^{\left(  1\right)  }\\
&  =-\left(  1+s\right)  \Upsilon^{\left(  1\right)  }+s\Upsilon^{\left(
2\right)  }+\left(  1-s\right)  \Upsilon^{\left(  3\right)  }+s\Upsilon
^{\left(  4\right)  },
\end{align*}
with
\[
a+b+c=1,\,\,\,0\leq t\leq1,\,\,s\in%
%TCIMACRO{\U{211d}}%
%BeginExpansion
\mathbb{R}%
%EndExpansion
\]
We use Proposition\thinspace\ref{prop:phi-max-extreme}.%

\[
G_{\Phi}^{\max}\left(  \Delta\right)  =\min_{\substack{s\in%
%TCIMACRO{\U{211d} }%
%BeginExpansion
\mathbb{R}
%EndExpansion
\\t\in\left[  0,\min\{a,c\}\right]  }}\left[  \frac{\left(  1+s\right)  ^{2}%
}{a-t}+\frac{s^{2}}{b+t}+\frac{\left(  1-s\right)  ^{2}}{c-t}+\frac{s^{2}}%
{t}\right]
\]
First, we optimize over $s$, which achieves minimum at
\[
s=\frac{\left(  a-c\right)  t\left(  t+b\right)  }{-t^{2}+2act+abc}.
\]
Hence,
\begin{align*}
G_{\Phi}^{\max}\left(  \Delta\right)   &  =\min_{t\in\left[  0,\min
\{a,c\}\right]  }\frac{2t+ab+bc}{-t^{2}+2act+abc}\allowbreak\\
&  =\min_{t\in\left[  0,\min\{a,c\}\right]  }\frac{2t+ab+bc}{\left(  \left(
ac+\sqrt{a^{2}c^{2}+abc}\right)  -t\right)  \left(  t-\left(  ac-\sqrt
{a^{2}c^{2}+abc}\right)  \right)  }%
\end{align*}
After some computation, one can verify
\[
ac+\sqrt{a^{2}c^{2}+abc}=ac+\sqrt{a^{2}c^{2}+ac\left(  1-a-c\right)  }\leq
\min\left\{  a,c\right\}  .
\]
Therefore, the function to be optimized is monotone increasing in the domain.
Hence, the minimum is achieved at $t=0$. Therefore,
\[
G_{\Phi}^{\max}\left(  \Delta\right)  =\frac{a+c}{ac}\allowbreak=\frac{1}%
{a}+\frac{1}{c}.
\]
Note that the optimal simulation uses three extreme points, $\Upsilon^{\left(
1\right)  }$, $\Upsilon^{\left(  2\right)  }$, and $\ \Upsilon^{\left(
3\right)  }$. It is not difficult to compute
\[
G_{\Phi}^{\min}\left(  \Delta\right)  =\max\left\{  \frac{1}{a}+\frac{1}%
{1-a},\frac{1}{c}+\frac{1}{1-c}\right\}  .
\]
Since $a+c\leq1$, $G_{\Phi}^{\max}\left(  \Delta\right)  \geq G_{\Phi}^{\min
}\left(  \Delta\right)  $. \ ("$=$" holds if and only if $a+c=1$.)

\end{document}